\documentclass[aps,physrev,twocolumn,groupedaddress]{revtex4-2}

\usepackage{amsmath}
\usepackage{graphicx}
\usepackage{gensymb}
\usepackage{xcolor}
\usepackage{hyperref}
\hypersetup{colorlinks=true,citecolor=violet, linkcolor=orange, filecolor=magenta, urlcolor=cyan,}

\begin{document}

\title{Capturing the three dimensional, nano-scale, pico-second dynamics of plasma mirrors with intense ultrashort laser wavefront measurement}


\author{Sk Rakeeb, Sagar Dam, Ameya Parab, Amit Lad, Yash M. Ved, G. Ravindra Kumar}
\thanks{grk@tifr.res.in}
\affiliation{Tata Institute of Fundamental Research, Mumbai, India 400005}


\date{\today}

\begin{abstract}
We present a direct measurement of the nanoscale dynamics of plasma mirrors using wavefront measurement techniques. This two-dimensional measurement, performed via pump-probe diagnostics, enables the reconstruction of the three-dimensional plasma mirror surface with nanometer axial, micrometer transverse, and femtosecond temporal resolution.

\end{abstract}


\maketitle

Recent developments in ultrahigh intensity laser technology open up possibilities for high energy density research \cite{drake2010high,kaw2017nonlinear} with critical applications in inertial confinement fusion (ICF) \cite{betti2016inertial}, laser-driven particle acceleration \cite{norreys2009laser}, high-brightness VUV and X-ray sources \cite{dromey2007bright}, attosecond optics \cite{thaury2007plasma}, tabletop ultrafast X-ray imaging \cite{gaffney2007imaging} and more. The temporal and spatial properties of femtosecond pulses are crucial parameters for advancement on all these fronts \cite{froula2018spatiotemporal}. While interacting with solid matter, high contrast pulses excite solid density matter ``instantaneously",  in contrast to a pulse with a strong picosecond wing and a nanosecond pedestal where the interaction takes place in expanding, lower-than-solid density plasma. After the interaction, the `hot', `dense' plasma starts expanding towards vacuum and also launches shock waves into the solid.  These extreme conditions of matter can resemble some exotic scenarios in intrastellar and planetary environments. Once the plasma is formed, laser interaction is crucially impacted by the critical density surface, which absorbs and reflects the light, thus referred to as the `plasma mirror'.

One of the key challenges in these experiments is the determination and control of the thin electron density gradient that emerges at the plasma-vacuum boundary\cite{he2015coherent}. High-order harmonic generation, conducted with a carefully tailored pre-pulse\cite{kahaly2013direct} , has provided direct insight into the transition toward the relativistic oscillating-mirror regime when the density gradient remains much smaller than the laser wavelength $(L/\lambda << 1)$. Additionally, plasmas with steeper density gradients  have led to electron acceleration reaching energies in the few-100 keV range\cite{bastiani1997experimental,mordovanakis2009quasimonoenergetic}. Short pulse ion acceleration experiments are also heavily affected by the density gradients. Therefore, it is crucial to examine the dynamics and evolution of these plasma mirrors in picosecond timescales.

Over the past few decades, a wide range of plasma diagnostic techniques have been implemented to study plasma dynamics. Doppler spectrometry \cite{mondal2010doppler, adak2015terahertz, jana2018probing} and velocity interferometry system for any reflector  (VISAR) \cite{barker1972laser,celliers1998accurate} have proven effective in measuring plasma expansion velocity. For precise phase-sensitive measurements, frequency domain interferometry (FDI) \cite{geindre1994frequency, dulat2022subpicosecond, dulat2024single} stands out as a strong candidate, offering an impressive resolution of $\lambda/2000$. However, its application to a plasma system is complex. A more accessible alternative is spatial domain interferometry (SDI) \cite{bocoum2015spatial}, which, is easier to set up but still shares the same limitation with other techniques—these diagnostics suffer from low spatial resolution, sometimes approximating an entire 3D plasma structure to a single point. Two-dimensional interferometric techniques can experience loss of interferometric fringes in overdense regions, making density measurements challenging. Similarly, shadowgraphy \cite{batani2019optical}, which is useful for visualizing density gradients, primarily provides qualitative rather than absolute electron or ion density values, limiting its effectiveness in precise plasma characterization.

In this paper, we demonstrate a new and easy-to-implement diagnostic to characterize the plasma mirror from a laser-solid interaction. This approach utilizes a wavefront measurement device to quantify wavefront modifications in a pump-probe diagnostic setup. By employing this method, the plasma surface can be reconstructed in three dimensions with high spatial resolution. Wavefront measurement techniques have previously been applied to determine laser-induced damage thresholds \cite{sozet2017sub,gallais2016time} and to measure electron density in gaseous, under-critical plasmas \cite{plateau2010wavefront,huijts2022waveform}. Notably, wavefront measurement has also become a standard tool in biology for precisely mapping cells and other microscopic structures at nanometric scales \cite{bon2009quadriwave,baffou2023wavefront}.

   \begin{figure}[h]
        \centering
        \includegraphics[width=1\linewidth]{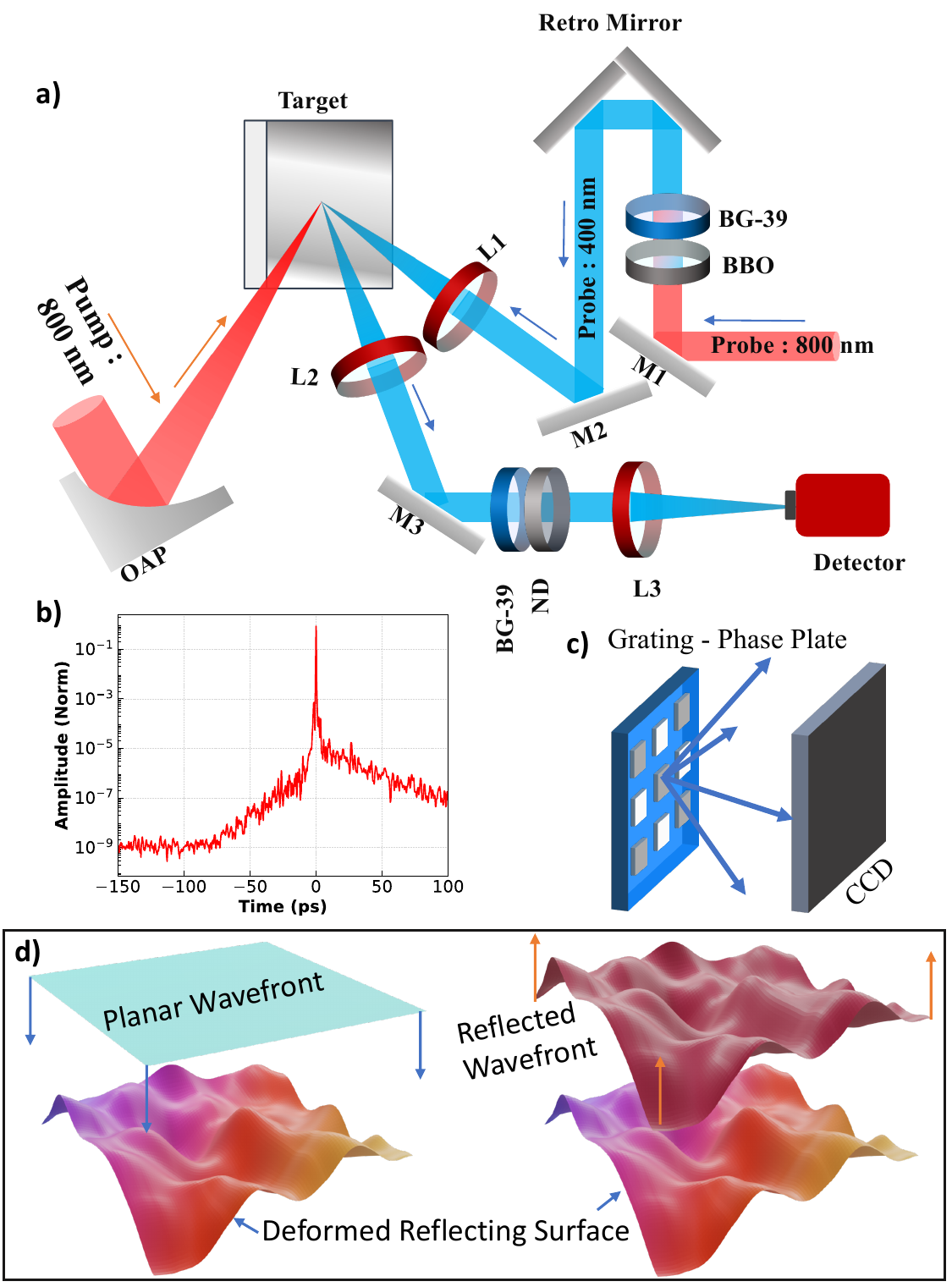}
        
        \caption{\textbf{a.} Experimental setup for the measurement. Pump and probe was matched specially and temporally on the target \textbf{b.} Measured temporal contrast of the laser \textbf{c.} Schematic of a QWLSI wavefront measuring device. Four replicas of the wavefront to be measured is created and the interferogram is recorded in the CCD \textbf{d.} Principle of the measurement. The reflected wavefront carries the information of the reflecting plasma surface }
        \label{exp_setup}
    \end{figure}
As shown in Figure~\ref{exp_setup}, a high-contrast ($10^{-9}$ at -100 ps), p-polarized Ti:Sapphire laser (30 fs, 800 nm) was used to generate overdense plasma from solid glass (BK7) targets. The laser was focused using an f/3 off-axis dielectric-coated parabola at an incidence angle of 45\degree, achieving a 7 $\mu m$ focal spot (FWHM) with an intensity of $2 \times 10^{17}$ W/cm\textsuperscript{2}.

A small fraction (5\%) of the main beam, referred to as the probe, extracted and temporally delayed relative to the original beam (the pump) was used to investigate the plasma's temporal evolution. The probe beam was frequency-doubled to its second harmonic (400 nm) using a Type-II BBO crystal, with BG-39 filters employed to remove residual 800 nm light. It was then focused onto the target at near-normal incidence (5\degree) producing a focal spot of $200 \mu m$.
The temporal delay with respect to the pump was controlled using a retro-reflector. The temporal resolution (200 fs) was primarily constrained by the temporal width of the upconverted probe. Spatial and temporal alignment of the pump and probe on the target plane was carefully achieved, as illustrated in Figure~\ref{exp_setup}a. After reflection, the probe was collected and imaged onto a PHASICS SID-4 wavefront sensor with a magnification of 10. 

The wavefront measurement system employed in this study is a high-definition device known as the QuadriWave Lateral Shearing Interferometer (QWLSI), originally developed by Primot et al. in 2000 \cite{primot2000extended, primot1995achromatic,velghe2005wave}. The QWLSI system consists of a two-dimensional transmission grating, a phase checkerboard and a CCD sensor. The pitch of the phase checkerboard is twice that of the grating pitch, as seen in Figure~\ref{exp_setup}c, where the dark squares on the grating phase plate contain half-wave plates, while the light squares are left blank. The addition of this phase checkerboard effectively cancels the \(0, \pm3\), and other even diffraction orders produced by the grating. The dominant contributions arise from the four first-order diffractions, which are tilted at an angle \(\theta \approx \lambda/d\) relative to the \(z\)-axis, where \(d\) is the grating pitch. These tilted replicas interfere with each other, generating an interferogram spots along the \(x\) and \(y\) directions.
The raw interferogram is subjected to a Fourier transform, after which the first-order components in the \(x\) and \(y\) directions are selectively masked and inverse Fourier transformed to retrieve the phase. However, the resulting phase images are wrapped, necessitating a phase unwrapping procedure to eliminate \(2\pi\) ambiguities and obtain a continuous phase distribution.
The addition of the phase checkerboard renders the wavefront sensor achromatic, which may seem counterintuitive \cite{chanteloup2005multiple}. Although our probe pulse has a spectral width of only 2 nm the device can also be used with broadband laser sources. However, this achromaticity does not imply that the interferogram quality is entirely unaffected by wavelength. In practice, using a QWLSI wavefront sensor at a wavelength other than its design wavelength leads to a reduction in interferogram contrast, which in turn decreases the sensitivity of the phase measurement.
The wavefront measurement device used in the experiment has a surface RMS accuracy of 5 nm and a high spatial resolution of 27 $\mu$m. Consequently, such devices are widely employed in adaptive optics for wavefront correction in high-intensity laser systems, where distortions arise from multiple amplification stages. Additionally, they are extensively utilized in phase microscopy for biological imaging \cite{baffou2023wavefront}.

The measurement is performed as follows: First, 20 shots are recorded without the interaction of the pump beam (achieved by physically blocking the pump beam with a shutter placed inside the vacuum chamber), and the average measurement is taken as the reference. Then, laser shots of the pump interacting with the target are recorded for a particular delay between the pump and probe, and the delay is varied. Multiple shots are taken to mitigate minor shot-to-shot jitter and reduce numerical errors.
The phase difference is determined by subtracting the phase of the probe without pump interaction from the phase of the probe with pump interaction.

   \begin{figure}[h]
        \centering
        \includegraphics[width=1\linewidth]{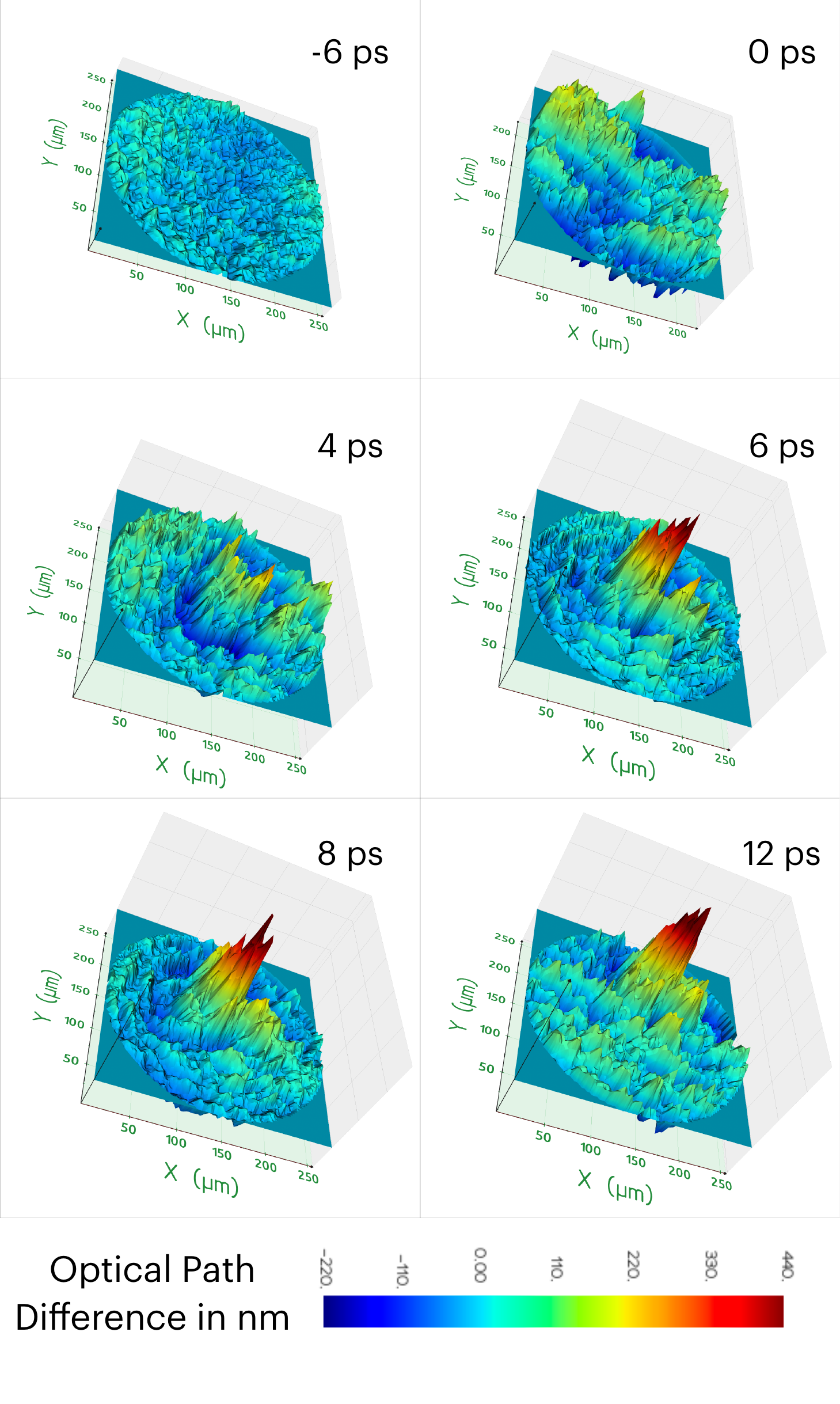}
        
        \caption{Measured wavefront at different time delay between pump and probe. Initially near $T_0$ (i.e. pump-probe delay 0 ps), the plasma is pushed inside and at later times, the plasma expands towards the vacuum due to the thermal expansion.}
        \label{wf}
    \end{figure}


The measured wavefronts at different time delays are presented in Figure~\ref{wf}, capture the high resolution 2- dimensional behavior of the plasma surface . When the low intensity probe arrives before the pump, the phase changes at negative delays remain insignificant as expected. At $T_0$, when the pump and probe reach the target simultaneously, the phase shift at the center is negative, indicating an inward displacement of the plasma surface due to the extreme pressure exerted by the pump pulse. During the pump interaction, the ponderomotive force displaces electrons and ions toward the lower-intensity region, forming a depression at the highest-intensity point. At later time delays, once the pump laser has dissipated, the plasma (assumed to be self-similar) begins to expand exponentially into the vacuum, generating a density gradient. To validate our findings, we compare the measured data with previous Doppler spectrometry results \cite{jana2018probing} obtained under similar experimental conditions. The central pixels of the measured wavefront were averaged, and the corresponding data were plotted alongside Doppler measurements in Figure~\ref{wf_doppler}. A sharp increase in plasma velocity occurs around 5 ps, coinciding with a rapid rise in the surface deformation. Beyond this point, the plasma expansion decelerates, leading to a more gradual evolution.

    \begin{figure}[htbp!]
        \centering
        \includegraphics[width=1\linewidth]{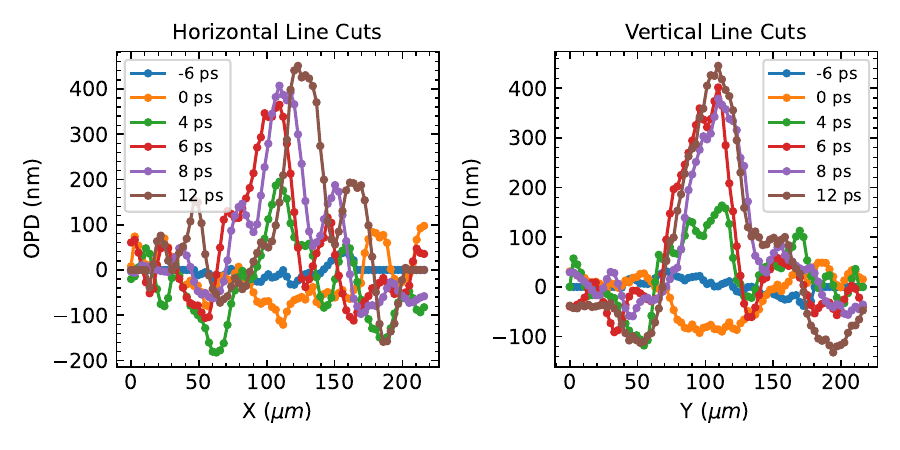}
        
        \caption{The horizontal and vertical line-cut data of the measured wavefront at different delays. At 0 ps, the wavefront exhibits clear negative values, indicating an inward displacement of the plasma surface. At later time delays, the plasma evolves like a rippled structure, where the central region expands outward while density depressions form at the sides.}
        \label{wf_doppler}
    \end{figure}

When analyzing phase changes to model the plasma surface, it is essential to account for two key factors: the motion of the critical surface and the plasma expansion in the underdense region.
The total optical path difference between the part of probe interacting with the uninteracted target surface and the part of probe interacting with the plasma is given by:

\begin{align}
    \Delta L &=  2\int_{\infty}^{x_c} n(x) dx - 2\int_{\infty}^{0} dx  \\
             &=  2\int_{\infty}^{x_c} \left( 1 - \frac{n_e}{n_c} \right)^{1/2} dx - 2\int_{\infty}^{x_c} dx  + 2x_c  \\
             &\approx 2\int_{\infty}^{x_c} \left( 1 - \frac{n_e}{2n_c} \right) dx - 2\int_{\infty}^{x_c} dx  + 2x_c  \\
             &= 2x_c + \int_{x_c}^{\infty} \frac{n_e}{n_c} dx
\end{align}

where \(x_c\) denotes the position of the critical surface from the target plane, $n$ represents the refractive index, $n_e$ and $n_c$ are the electron and critical density respectively. Assuming an exponential density profile of the form: $n_e = n_c exp[-(x - x_c)/l]$, the optical path difference:


\begin{align}
    \Delta L &= 2x_c + \int_{x_c}^{\infty} e^{-\frac{x - x_c}{l}}  dx\\
             &= 2x_c + l
\end{align}

where \(l\) represents the density scale length. For a BK7 glass target, the density is approximately \(150 n_c\), leading to \(x_c = l \ln(150)\). This results in \(l \approx x_c/5\), which simplifies the expression for critical surface position to: $\Delta L =2.2x_c$. Thus the measured path difference can be directly related to the position of the critical surface. This approach provides a simplified analytical framework for estimating the position of the critical surface. However, alternative models can be employed depending on the specific experimental conditions to achieve a more accurate mapping of the critical surface dynamics.

    \begin{figure}[htbp!]
        \centering
        \includegraphics[width=1\linewidth]{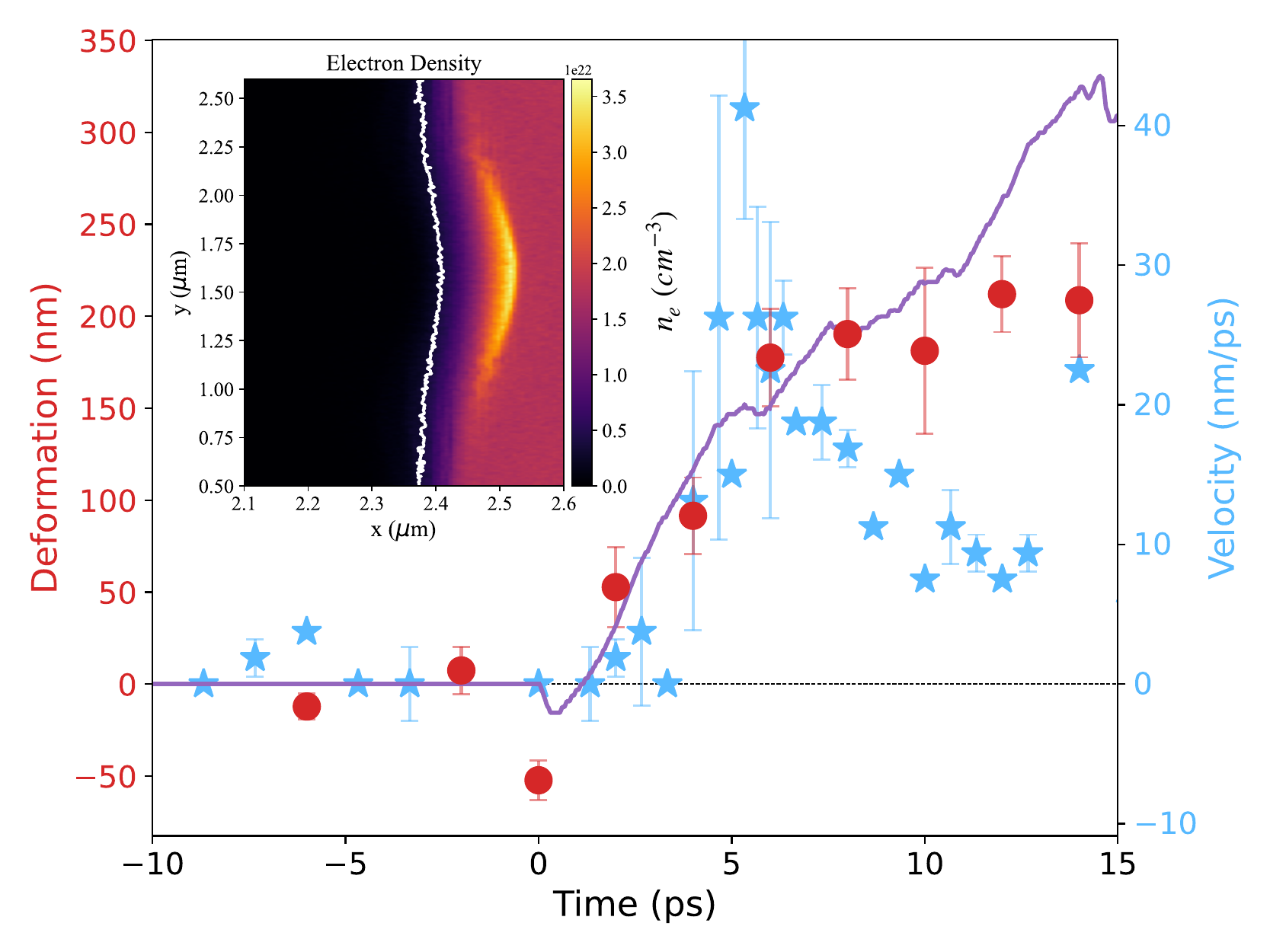}
        
        \caption{The deformation of the critical surface as a function of different pump-probe deyays(red circles). For comparison, velocity data form Jana et al.\cite{jana2018probing} (blue stars)is shown. The solid line shows the 1D simulation results. The inset illustrates the ponderomotive dent observed in the 2D simulation. The white line indicates the position of the 400 nm critical surface at 250 fs.}
        \label{wf_doppler}
    \end{figure}



Particle-in-cell (PIC) simulations were performed to gain deeper insights into the underlying plasma dynamics. 1D3V simulations were conducted using the SMILEI PIC code \cite{derouillat2018smilei} with a box size of 6.4 $\mu$m and a spatial resolution of 3 nm, containing a plasma slab extending 2.4 $\mu$m. The simulation was initialized with approximately $4 \times 10^5$ particles. A Gaussian laser pulse with a temporal FWHM of 27 fs and intensity similar to the experiment was incident from the $x_{\text{min}}$ boundary of the simulation box. The hydrodynamic expansion of the plasma was studied up to 15 ps with a temporal resolution of 7 attoseconds. The ions were modeled with a mass 26 times that of protons and an atomic number of 13, closely matching the average atomic number of BK7 glass (13.5). Ionization and collisional effects were incorporated to ensure a realistic simulation closer to experimental conditions. Collisional calculations were performed every 20 time steps (140 attoseconds) to accurately capture the plasma evolution. The simulation results are presented in Figure~\ref{wf_doppler}.

Additionally, 2D PIC simulations were initiated with electron-proton plasma including 17 million particles within a computational domain of $3.2 \mu$m $\times\ 3.2 \mu$m, a spatial resolution of 128, and other parameters identical to those of the 1D simulation. The plasma dynamics were recorded over a duration of 2 ps. The inset of Figure~\ref{wf_doppler} illustrates the simulated ponderomotive dent at 250 fs. Notably, the simulation results exhibit a trend similar to the experimental observations, further corroborating our measurements.


In conclusion, we have presented a direct method for measuring two-dimensional plasma dynamics on picosecond timescales by analyzing the wavefront of the probe beam. With the rise of high-power ultrashort laser pulses and their applications, this technique offers a direct measure of the instantaneous critical surface and thereby providing detailed insight into plasma conditions. This is particularly valuable for optimizing plasma parameters for specific studies or applications. Additionally, numerical simulations have been performed, strongly supporting the validity of our measurements.


%
%

\begin{acknowledgments}
G.R.K. acknowledges support from the J.C. Bose Fellowship grant JBR/2020/000039. SR acknowledges the TIFR cluster facility for providing access to the HPC to perform the simulations. SR thanks Prof. Laszlo Veisz for the insightful discussion.
\end{acknowledgments}

\bibliography{references}
\end{document}